\documentclass[reprint,aps,prl,superscriptaddress,longbibliography]{revtex4-1}
\usepackage[latin9]{inputenc}
\usepackage{color}
\usepackage{times}
\usepackage{amsmath}
\usepackage{amssymb}
\usepackage{stmaryrd}
\usepackage{graphicx}
\usepackage[unicode=true,
 bookmarks=true,bookmarksnumbered=false,bookmarksopen=false,
 breaklinks=false,pdfborder={0 0 1},backref=false,colorlinks=true]
 {hyperref}
\hypersetup{
 linkcolor=magenta, urlcolor=blue, citecolor=blue, pdfstartview={FitH}, hyperfootnotes=true, unicode=true}
 
\makeatletter
\@ifundefined{textcolor}{}
{%
 \definecolor{BLACK}{gray}{0}
 \definecolor{WHITE}{gray}{1}
 \definecolor{RED}{rgb}{1,0,0}
 \definecolor{GREEN}{rgb}{0,1,0}
 \definecolor{BLUE}{rgb}{0,0,1}
 \definecolor{CYAN}{cmyk}{1,0,0,0}
 \definecolor{MAGENTA}{cmyk}{0,1,0,0}
 \definecolor{YELLOW}{cmyk}{0,0,1,0}
}

\newcommand{\doublewidetilde}[1]{{%
  \mathpalette\double@widetilde{#1}%
}}
\newcommand{\double@widetilde}[2]{%
  \sbox\z@{$\m@th#1\widetilde{#2}$}%
  \ht\z@=.9\ht\z@
  \widetilde{\box\z@}%
}

\definecolor{darkgreen}{rgb}{0.0, 0.6, 0.13}

\usepackage{amsfonts}\usepackage{tabularx}\usepackage{dcolumn}\usepackage{bm}\usepackage{graphicx}\usepackage{epstopdf}

\hypersetup{urlcolor=blue}

\usepackage{tensor}
\usepackage{braket}

\usepackage[capitalise,compress]{cleveref}
\crefname{section}{Sec.}{Secs.}
\Crefname{section}{Section}{Sections}
\crefrangelabelformat{equation}{\textup{(#3#1#4)}--\textup{(#5#2#6)}}

\makeatother
\begin{document}

\title{Confined Quasiparticle Dynamics in Long-Range Interacting Quantum Spin Chains
}

\author{Fangli Liu}
\affiliation{Joint Quantum Institute, NIST/University of Maryland, College Park, MD 20742, USA}

\author{Rex Lundgren}
\affiliation{Joint Quantum Institute, NIST/University of Maryland, College Park, MD 20742, USA}

\author{Paraj Titum}
\affiliation{Joint Quantum Institute, NIST/University of Maryland, College Park, MD 20742, USA}
\affiliation{Joint Center for Quantum Information and Computer Science, NIST/University of Maryland, College Park, MD 20742, USA}

\author{Guido Pagano}
\affiliation{Joint Quantum Institute, NIST/University of Maryland, College Park, MD 20742, USA}

\author{Jiehang Zhang}
\affiliation{Joint Quantum Institute, NIST/University of Maryland, College Park, MD 20742, USA}

\author{\\Christopher Monroe}

\affiliation{Joint Quantum Institute, NIST/University of Maryland, College Park, MD 20742, USA}

\affiliation{Joint Center for Quantum Information and Computer Science, NIST/University of Maryland, College Park, MD 20742, USA}

\author{Alexey V. Gorshkov}
\affiliation{Joint Quantum Institute, NIST/University of Maryland, College Park, MD 20742, USA}
\affiliation{Joint Center for Quantum Information and Computer Science, NIST/University of Maryland, College Park, MD 20742, USA}

\begin{abstract}

We study the quasiparticle excitation and quench dynamics of the one-dimensional transverse-field Ising model with power-law ($1/r^{\alpha}$) interactions. We find that long-range interactions give rise to a confining potential, which couples pairs of domain walls (kinks) into bound quasiparticles, analogous to mesonic bound states in high-energy physics. We show that these quasiparticles have signatures in the dynamics of order parameters following a global quench and the Fourier spectrum of these order parameters can be expolited as a direct probe of the masses of the confined quasiparticles.  We introduce a two-kink model to qualitatively explain the phenomenon of long-range-interaction-induced confinement, and  to quantitatively predict  the masses of the bound quasiparticles.  Furthermore, we illustrate that these quasiparticle states can lead to slow thermalization of one-point observables for certain initial states.  Our work  is readily applicable to current trapped-ion experiments.
\end{abstract}

\pacs{}

\maketitle

Long-range interacting quantum systems occur naturally in numerous quantum simulators \cite{Saffman10, Schau12, Islam13, Dolde13, Lu12, Child06, Weber10, Gopala11, Douglas15, Hung16}. A paradigmatic model considers interactions decaying with distance $r$ as a power law $1/r^\alpha$. This describes the interaction term in trapped-ion spin systems~\citep{Islam13, Britton12, Richer14, Jurcevic14, Smith16, Zhang17}, polar molecules \cite{Ni08, Ni10, Chotia12, Molony14}, magnetic atoms \cite{Lu12, Aikawa, Bala09}, and Rydberg atoms \cite{Saffman10, Schau12, Beguin13, Dudin12}. One remarkable consequence of long-range interactions is the breakdown of locality, where quantum information, bounded by linear `light cones' in short-range interacting systems \cite{Lieb72}, can propagate super-ballistically or even instantaneously \cite{Gong14, Foss14, Hauke13, Schach13, Vander18, Buy16}. Lieb-Robinson linear light cones have been generalized to logarithmic 
and polynomial 
light cones  for long-range interacting systems \cite{Hasting06, Gong14, Foss14}, and non-local propagation of quantum correlations in one-dimensional (1D) spin chains has been observed in trapped-ion experiments \cite{Richer14, Jurcevic14}.  Moreover, 1D long-range interacting quantum spin chains can host novel physics that is absent in their short-range counterparts, such as continuous symmetry breaking \cite{Maghrebi17, Mermin66}.

 More recently, it has been shown that  confinement--which has origins in high-energy physics--has dramatic signatures in the quantum quench dynamics of short-range interacting spin chains \cite{Kormos17}. Owing to confinement, quarks cannot be directly observed in nature as they form mesons and baryons due to strong interactions~\cite{Green11, Vander12}.  An archetypal model with  analogous confinement effects in  quantum many-body systems is the 1D short-range interacting Ising model with both transverse and longitudinal fields \cite{Coldea10, Rutkev08, Morris14, Kjall, Delfino14, Delfino17}. For a vanishing longitudinal field, domain-wall quasiparticles propagate freely and map out light-cone spreading of quantum information \cite{Calabrese11, Delfino14, Delfino17, Delfino18}.  As first proposed by McCoy and Wu~\cite{Mccoy78, Mccoy68, Mccoy12}, a non-zero longitudinal field induces an attractive linear potential between two domain walls and confines them into mesonic bound quasiparticles. Recently,  Kormos et al.~investigated the effect of these bound states on quench dynamics and showed that the non-equilibrium dynamics can be used to probe confinement \cite{Kormos17}. 

In this work, we study the non-equilibrium dynamics of the long-range interacting transverse-field Ising model \emph{without a longitudinal field} after a global quantum quench. We find that long-range interactions introduce an effective attractive force between a pair of domain walls, thus confining them into a bound state, analogous to the meson in high-energy physics. 
We calculate time-dependent order parameters and connected correlation functions, both of which feature clear signatures of bound quasiparticle excitations \cite{Delfino14, Delfino17}.  The masses of these bound quasiparticles--the energy gaps relative to the ground state--can be directly extracted from the Fourier spectrum of time-dependent order parameters \cite{Kormos17, Delfino14, Delfino17}. We introduce a two-kink model, i.e.\ we restrict spin configurations to states with only a pair of domain walls (kinks), 
to explicitly show that the confining potential comes from long-range interactions. The two-kink model also gives good predictions for the quasiparticles' masses and their dispersion relations. Furthermore, we study the effect of confined quasiparticles on the thermalization of different initial states \cite{Delfino14, Delfino17}. We find that for certain initial states, one-point observables exhibit slow thermalization \cite{Banuls11, Lin17, Delfino14, Delfino17}, which might help protect ordered phases in the prethermal region \cite{Zhang17Time, Hess17, Machado17}.

We note that our study is in agreement with the general mechanism of global quantum quenches, first formulated in Refs. \cite{Delfino14, Delfino17, Delfino18} for short-range interacting systems, and demonstrates that the general theory developed in Refs.~\cite{Delfino14, Delfino17, Delfino18} holds for systems with long-range interactions. Our work is  well within the reach of current trapped-ion experiments~\cite{Zhang17} and other atomic, molecular, and optical (AMO) experimental platforms~\cite{Saffman10,Douglas15, Bernien17}.
\begin{figure}
  \centering\includegraphics[width=0.485\textwidth, height=6.5cm]{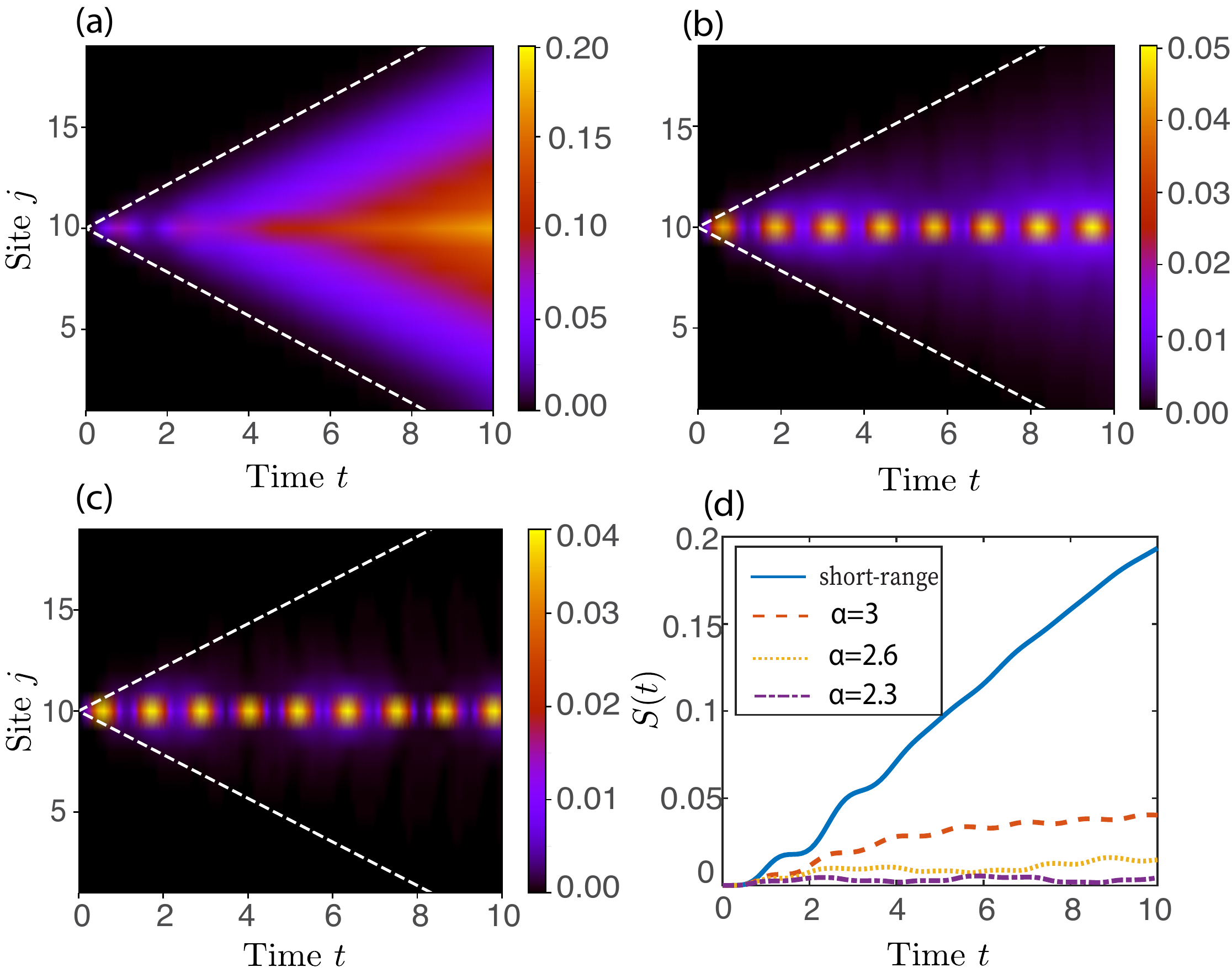}
  \caption{(color online) (a)-(c) $ \langle \sigma^z_j \sigma^z_k\rangle_c$,  and (d) $S_A(t)$ versus $t$ after a quantum quench with initial state $|\Psi_0\rangle$. $L= 19$, $k=10$, and $B=0.27$. (a) Short-range interacting case ($\alpha\rightarrow\infty$), (b) $\alpha= 2.6$, (c) $\alpha= 2.3$. The dashed white lines illustrate the maximal velocity, $4B$,  of freely propagating domain walls in the short-range interacting case~\cite{Kormos17}. (d) $S_A(t)$ for various $\alpha$.
  }
  \label{fig1}
\end{figure}

{\it The model.---} Let us consider a quantum spin chain with long-range interactions, described by the following Hamiltonian,
\begin{equation}
H= - \sum_{i<j}^L \frac{J}{r_{ij} ^{\alpha}} \sigma^z_i \sigma^z_j  - B \sum_{i=1}^L \sigma_i^x,
\label{Ham}
\end{equation}
where $\sigma_i^\mu$ are the Pauli matrices on site $i$, $L$ is the system size, $r_{ij}$ is the distance between sites $i$ and $j$ (nearest-neighbor spacing is assumed to be equal to 1),  $J$ sets the overall energy scale (taken to be 1 without loss of generality), $B$ is a global transverse magnetic field, and $\alpha$ describes the power-law decay of long-range interactions. In this work, we consider periodic  boundary conditions unless otherwise specified ($r_{ij}$ is then the shortest distance between sites $i$ and $j$). 

In the nearest-neighbor interacting limit ($\alpha\rightarrow \infty$), $H$ is exactly solvable via a Jordan-Wigner mapping to spinless fermions. It exhibits a second-order phase transition at $B=1$, which separates the ferromagnetic and paramagnetic phases \cite{Sachdev11}. The phase transition persists if one turns on long-range interactions; however, the critical value of $B$ increases \cite{Koffel12,Knap13, Fey16, Fey18}.  In trapped-ion experiments, the range of the power-law exponent can be tuned within $0<\alpha<3$  by changing the detuning of the applied optical fields from phonon sidebands. 
We restrict the numerics to $\alpha>1$ in order to ensure a well-behaved thermodynamic limit (the case of $\alpha\in [0, 1]$ will be briefly discussed later). Several experiments have investigated the real-time dynamics of the above model (or closely related models), including  dynamical phase transitions \cite{Zhang17, Jurcevic17}, the non-local propagation of correlations \cite{Richer14, Jurcevic14}, the time-crystal phase \cite{Zhang17Time}, and many-body localization \cite{Smith16}.


{\it Quench dynamics.---} Let us first study the quench dynamics of the above model. 
We focus on a simple initial state with all spins polarized in the $z$ direction, $\ket{\Psi_0}= \ket{...\uparrow \uparrow \uparrow ...} $, which can be easily prepared in trapped-ion experiments  \cite{Zhang17}. The system is  allowed to evolve under the Hamiltonian \eqref{Ham}.  This is equivalent to a global quantum quench from zero to finite $B$ \cite{Delfino14, Delfino17, Zhang17}. In order to explore the physics of domain walls, we focus on quantum quenches within the ferromagnetic phase \cite{Calabrese11, Calabrese06}. Finally, while we have chosen a spin-polarized initial state,  confinement persists when the initial state is chosen as the ground state of Eq.~\eqref{Ham} with  $B$ in  the ferromagnetic region.

\begin{figure}
  \centering\includegraphics[width=0.5\textwidth, height=6.35cm]{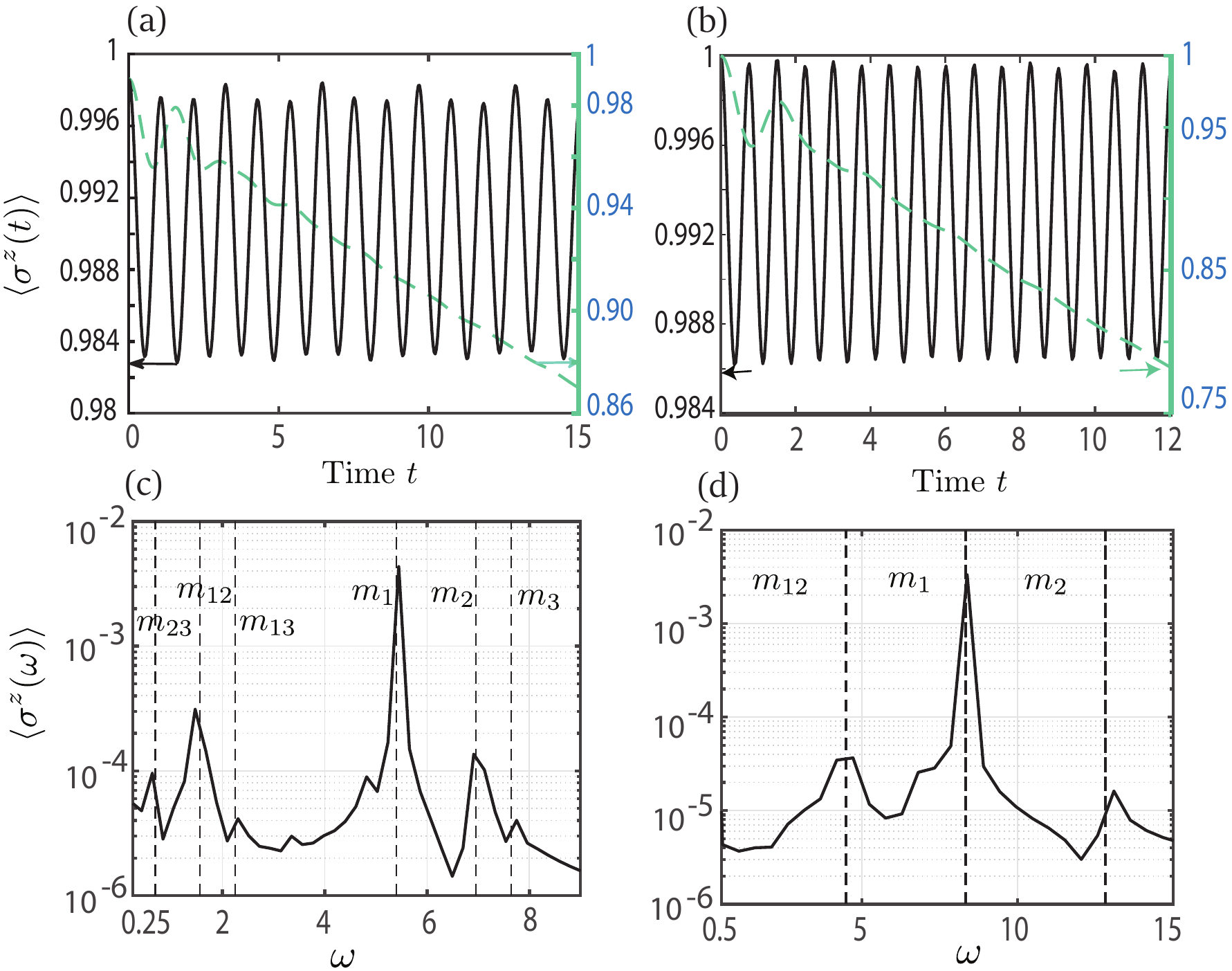}
  \caption{(color online)  (a)-(b) $\langle\sigma^z(t)\rangle$ (black line) versus time after quenching to (a) $\alpha= 2.3, B=0.27 $, (b) $\alpha =1.4, B=0.35$ for $L=20$. The dashed green lines show the decay of $\langle\sigma^z(t)\rangle$ for the short-range model with the same $B$. (c)-(d) Fourier spectrum of $\langle\sigma^z(t)\rangle$  for the long-range case in (a) and (b),  respectively. The largest time for the Fourier transform is up to $t=30$ and $12$ for (c) and (d), respectively. The parameters in (b,d) are accessible in current trapped-ion experiments~\cite{Zhang17}. The dashed lines show the mesonic masses ($m_i$) and their differences ($m_{ij}\equiv m_j-m_i$) calculated using the two-kink model.}
  \label{fig2}
\end{figure}

We use the Krylov-space method to simulate the quench dynamics of our system~\cite{Luitz17,Nauts83}. Figs.~\ref{fig1}(a)-(c) show the equal-time connected correlation functions, $\langle \sigma^z_j(t) \sigma^z_k(t)\rangle_c= \langle \sigma^z_j(t) \sigma^z_k(t)\rangle- \langle \sigma^z_j(t) \rangle \langle  \sigma^z_k(t)\rangle $, after the sudden quench (we take $k$ to be the central lattice site). In the short-range interacting limit [Fig.\ \ref{fig1}(a)], we recover the exactly solvable case, where correlations spread with a velocity ($4B$) equal to twice the maximal speed of free domain-walls ~\cite{Calabrese11, Kormos17, Delfino18}.  Increasing the Ising interaction range (decreasing $\alpha$) strongly suppresses the magnitude of $\langle \sigma^z_j(t) \sigma^z_k(t)\rangle_c$, as shown in Figs.~\ref{fig1}(b) and \ref{fig1}(c). One can also see the oscillatory behaviour of correlations in Figs.~\ref{fig1}(b) and \ref{fig1}(c), similar to that of Ref.~\cite{Kormos17}. However, we emphasize that the light-cone spreading of correlations is always present \cite{Kormos17, Delfino18}, though it may have a different velocity depending on the quasiparticles in the system \cite{Delfino18}.  The actual extent of the light cone becomes clearer by zooming in on the `black' regions of Figs.~\ref{fig1}(b) and \ref{fig1}(c) (see Supplemental Material for details \cite{supp}).  This result is in agreement with the general mechanism of global quantum quenches first derived in Ref. \cite{Delfino18}.

The propagating quasiparticles produced by the quench map out the light-cone spreading of correlations \cite{Delfino18, supp} and also lead to the growth of entanglement entropy \cite{ Kormos17}. In Fig.~\ref{fig1}(d), we plot the growth of entanglement entropy,  $S_A(t)= - \mathrm{Tr}[\rho_A(t)\mathrm{ln}(\rho_A(t))]$, where $\rho_A(t)$ is the reduced density matrix of one half of the chain, for various $\alpha$.  As one can see, the entanglement entropy growth for smaller $\alpha$ is much slower than the short-range case (linear growth). This is because there are less propagating quasiparticles for longer-range interactions, i.e. most quasiparticles produced by the quench have zero momentum \cite{ Kormos17}.



We plot time-dependent order parameters $\langle\sigma^z(t)\rangle= \frac{1}{L} \sum_i \langle \sigma^z_i(t)\rangle$ in Figs.~\ref{fig2}(a) and (b) \footnote{For Fig.~\ref{fig2}(b), we use parameters and probing time relevant to current trapped-ion experiments \cite{Zhang17,Richer14, Jurcevic14}}. Different from the rapid exponential decay of  the magnetization for the short-range case, $\langle\sigma^z(t)\rangle$ exhibits periodic oscillations with almost no decay \cite{Delfino14, Delfino17, Zau17, Jad171, Jad172} in the time window we are showing.  We emphasize that the qualitative change in dynamics is caused by the long-range interactions, not by an additional longitudinal field as in the short-range interacting case \cite{Delfino14, Delfino17, Kormos17}.  The Fourier spectrum ~\footnote{The time step used in Fig.~\ref{fig2} is 0.05.} of $\langle\sigma^z(t)\rangle$ illustrates that the oscillations are 
associated with multiple different frequencies (Figs.~\ref{fig2}(c) and \ref{fig2}(d)).  As we will see, these frequencies are in good agreement with the masses (and their differences) of quasiparticles \cite{Delfino14, Delfino17}.

\begin{figure}
  \centering\includegraphics[width=0.48\textwidth, height=8.9cm]{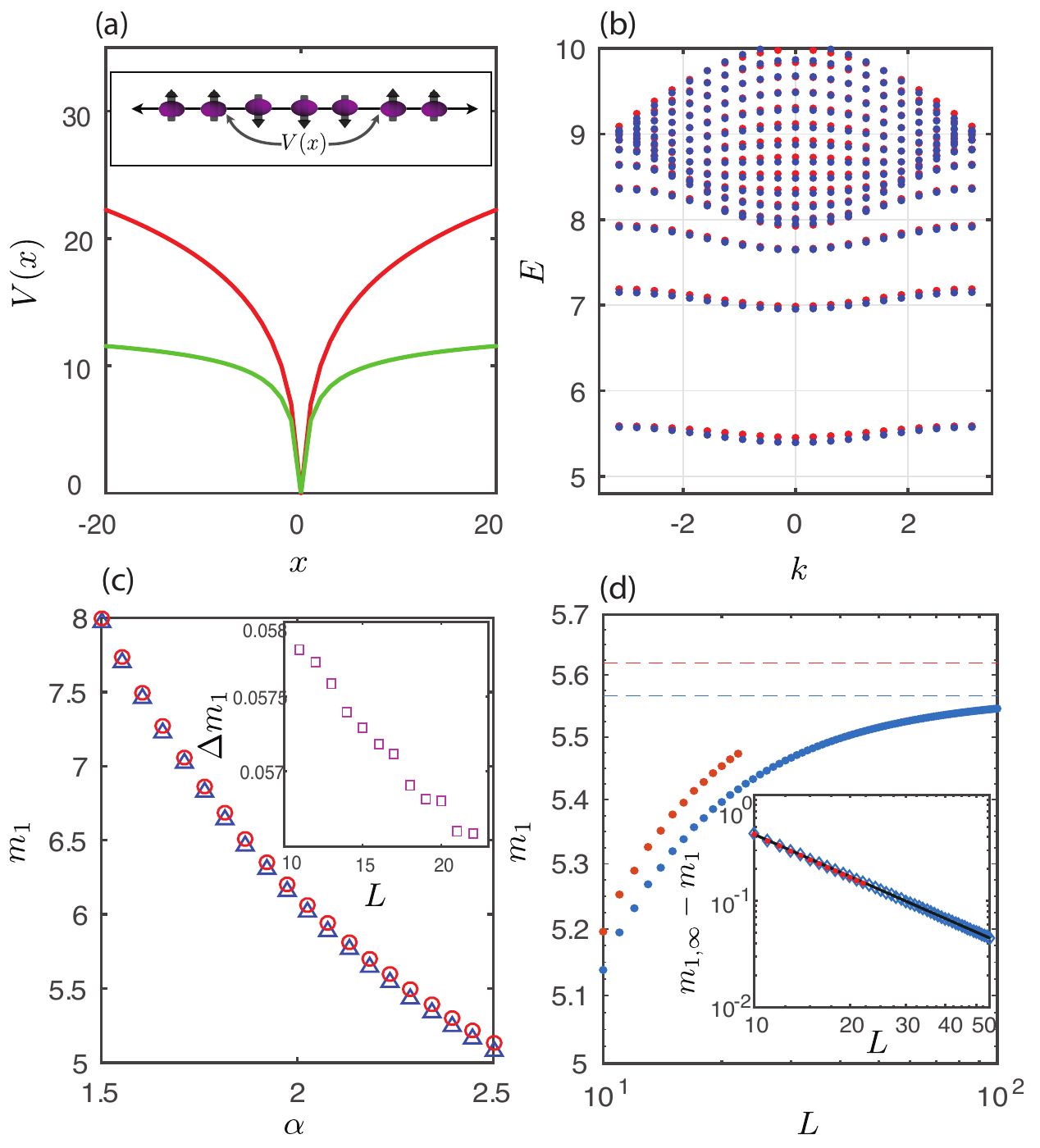}
  \caption{(color online) (a) Potential energy as a function of distance $x$ between the two domain walls ($n= \vert x \vert$). Red line: $\alpha=1.9$; green line: $\alpha=2.3$. Inset: typical spin configuration of two-domain-wall states. (b)-(d) Comparison of two-kink model (blue markers) and ED results (red markers) (b) Energy spectrum. Parameters: $\alpha=2.3, B=0.27$, $L=20$.  (c) $m_1$ versus $\alpha$ with parameters $B=0.27$ and $L=22$. Inset: Difference of $m_1$ between the two methods, $\Delta m_1$, versus $L$. (d) $m_1$ versus $L$, with the same parameters as in (b). The dashed lines are $m_{1,\infty}$. The inset shows $m_{1,\infty}-m_1$ versus system size. The black line shows the fitting of the two-kink model's data to $(1/L)^{\beta}$, with $\beta=1.315$. ED data has similar scaling with $\beta=1.34$. $m_{1, \infty}$ is chosen as 5.56 (5.62) for the two-kink model (ED). 
  }
\label{fig3}
\end{figure}

{\it Two-kink model and bound states.---} To understand the quasiparticles in our system, we use a two-kink model to perturbatively study the low-energy excitations of Eq.~\eqref{Ham}. The two-kink model has been used to phenomenologically study the confinement of excitations in short-range interacting quasi-1D compounds~\citep{Coldea10, Rutkevi10}. The idea is to 
restrict  the Hilbert-space to two domain-wall states [see inset of Fig.~\ref{fig3}(a)], where regions of different magnetization are separated by the two domain walls. The projected model is expected to work well when $B$ is much smaller than $J$\cite{Kormos17}.   

The Hilbert space of the projected model is spanned by states of $n$ down-spins (clustered together) which we represent as: $\ket{j, n}= \ket{...\uparrow\uparrow \downarrow_j\downarrow...\downarrow\downarrow_{(j+n-1)} \uparrow\uparrow...}$, where $j$ is the starting position of the cluster. The projected Hamiltonian, $\mathcal{H}= \mathcal{P} H \mathcal{P}$, where $\mathcal{P}$ is the projection operator to the two-domain-wall subspace, acts on $\ket{j, n}$ as follows,
\begin{equation}
\begin{split}
\mathcal{H}\ket{j, n}=& V(n)\ket{j, n}- B [\ket{j, n+1}+ \ket{j, n-1} \\ &+\ket{j+1, n+1}+ \ket{j-1, n+1}].
\end{split}
\end{equation}
Here, we have defined the potential energy 
as $V(n)=\langle j, n| \mathcal{H}| j,n\rangle-\langle\Psi_0|\mathcal{H}|\Psi_0\rangle$.  For our translational invariant system, the momentum $k$ is a good quantum number, and $\mathcal{H}$ is diagonal in the momentum basis. Fourier transforming the two-domain-wall state, $\ket{k, n} = \frac{1}{\sqrt{L}} \sum_{j=1}^{L} $exp$(-ik j-ikn/2) \ket{j, n}$, gives
\begin{equation}
\begin{split}
\mathcal{H}=\sum_{k,n} &V(n)|k,n\rangle\langle k,n|-2B\cos{\frac{k}{2}}|k,n\rangle\langle k,n+1| \\ &-2B\cos{\frac{k}{2}}|k,n\rangle\langle k,n-1|.
\end{split}
\label{eigen}
\end{equation}
For an infinitely large system, the potential energy of a two-domain-wall spin configuration is
\begin{align}
V(n)=
4n  \zeta (\alpha) J- 4J \sum_{1\leq l <n } \sum_{1\leq r \leq l} \frac{1}{r^{\alpha}},
\label{potential}
\end{align}
where $\zeta (\alpha)= \sum_{z=1}^{\infty} \frac{1}{z^{\alpha}}$ denotes the Riemann zeta function. As plotted in Fig.~\ref{fig3}(a), $V(n)$ \textit{increases} with the distance between domain walls.  For the short-range model studied by Kormos \emph{et} \emph{al}.~\cite{Kormos17}, the confining potential is due to an additional on-site longitudinal magnetic field. In our case, the confining potential is intrinsically generated  by the long-range interactions.

The picture now becomes clear: the long-range Ising interaction gives rise to an effective potential, which \textit{increases} with separation between domain walls, while the transverse magnetic field acts as kinetic energy for the two domain walls (increasing or decreasing the size of the cluster). Therefore, a pair of domain walls, each of which is free  quasiparticle in the short-range interacting limit, become bounded together when $\alpha$ decreases.  Note that $V(n)$ has an upper bound when $\alpha>2$, as illustrated in Fig.~\ref{fig3}(a) (see Supplemental Material \cite{supp}). This indicates that the lower part of the energy spectrum [obtained by diagonalizing Eq.~\eqref{eigen}] is composed of domain-wall bound states, while above some energy threshold, we have a continuum of states [Fig.~\ref{fig3}(b)]. For $\alpha \leq 2$, however, all excitations within the two-kink model are bound quasiparticles, as the confining potential $V(n)$  become unbounded when $n \rightarrow \infty$ \cite{supp}.  This is in contrast with finite-range interacting  models, where the potential  becomes \emph{flat} for $n$ greater than the interaction range. In other words, for finite-range interacting systems two domain walls will behave like freely propagating particles if the domain size of the initial state exceeds the interaction range. 


Fig.~\ref{fig3}(b) shows the energy spectrum calculated by the two-kink model (blue dots) and  exact diagonalization (ED) of the full Hilbert space (red dots).  As one can see, the energy spectrum agrees  well for the two methods, demonstrating that low-energy excitations are dominated by two-domain-wall states. The bound states' masses~\footnote{The energy difference between (bounded) excited states at $k=0$ and the ground state.} and dispersion relations can be simply read out from the energy spectrum. Moreover, the Fourier frequencies of $\langle\sigma^z(t)\rangle$ [Figs.~\ref{fig2}(c) and \ref{fig2}(d)] are compatible, to high accuracy, with the masses of the bound states (and their differences) calculated using the two-kink model \cite{Delfino14, Delfino17}.  This demonstrates that the quench dynamics of the long-range interacting model is indeed dominated by confined domain walls.


We compare the smallest bound state mass, $m_1$, as a function of $\alpha$ calculated using the two-kink model (blue) and ED (red) in Fig.~\ref{fig3}(c). For a large range of $\alpha$, we see excellent agreement between the two methods, and the numerical difference does not increase for larger $L$ [inset of Fig.~\ref{fig3}(c)]. The masses increase with $L$ as longer chains have more interaction terms~\cite{supp}. However, $V(n)$ is finite (for finite $n$) in the thermodynamic limit, since the Riemann zeta function converges for $\alpha>1$ \cite{Clair00}. This leads to finite masses, even for an infinite system when $\alpha>1$ (see Supplemental Material~\cite{supp}). Fig.~\ref{fig3}(d) shows the system-size dependence of $m_1$. The mass calculated from the two-kink model indeed exhibits convergence in the thermodynamic limit. For the two-kink model, the difference between $m_1$ and its thermodynamic value, $m_{1, \infty}$, scales as $(1/L)^{\beta}$, with $\beta \approx \alpha -1$ ~\cite{supp}, as shown in the inset. While we cannot verify convergence using ED, we do observe similar scaling of $m_1$ [inset of Fig.~\ref{fig3}(d)]. For $0\leq\alpha\leq 1$, $V(n)$ becomes infinite, even for finite $n$, and thus the bound states have  infinite energy (as the Riemann zeta function diverges for $0\leq\alpha\leq 1$~\cite{Clair00}), consistent with the results of Ref.~\cite{Santos16}.

{\it Strong and weak thermalization.---}   For the quenches we have considered, both the order parameter decay and entanglement entropy growth are quite slow (Fig.~\ref{fig1}). This movitates us to study thermalization in our long-range model. Previous studies of the short-range Ising model have observed rapid (strong) or slow (weak) thermalization of one-point functions for  different initial states~\cite{Delfino14, Delfino17, Banuls11, Lin17, James184, Mazza186, Rak16, Robin188}.  As first shown in Ref.~\cite{Delfino14}, undamped oscillations (weak thermalization) of one-point observable occurs within an intermediate time window when the matrix element between the initial state and the quasiparticle state of the quench operator and of the observable are both non-zero \cite{Delfino14, Delfino17}. Rapid decay occurs when this condition is not satisfied. Numerical results consistent with this finding have been observed in Refs. \cite{Kormos17, Rak16, Delfino17, Banuls11, Lin17}.  Here, we illustrate that these two distinct thermalization behaviors also occur in the long-range Ising model and that slow thermalization can arise
when the quasiparticles are the result of confinement \cite{Rak16, Delfino17, Kormos17}. 
\begin{figure}
  \centering\includegraphics[width=0.46\textwidth, height=3.7cm]{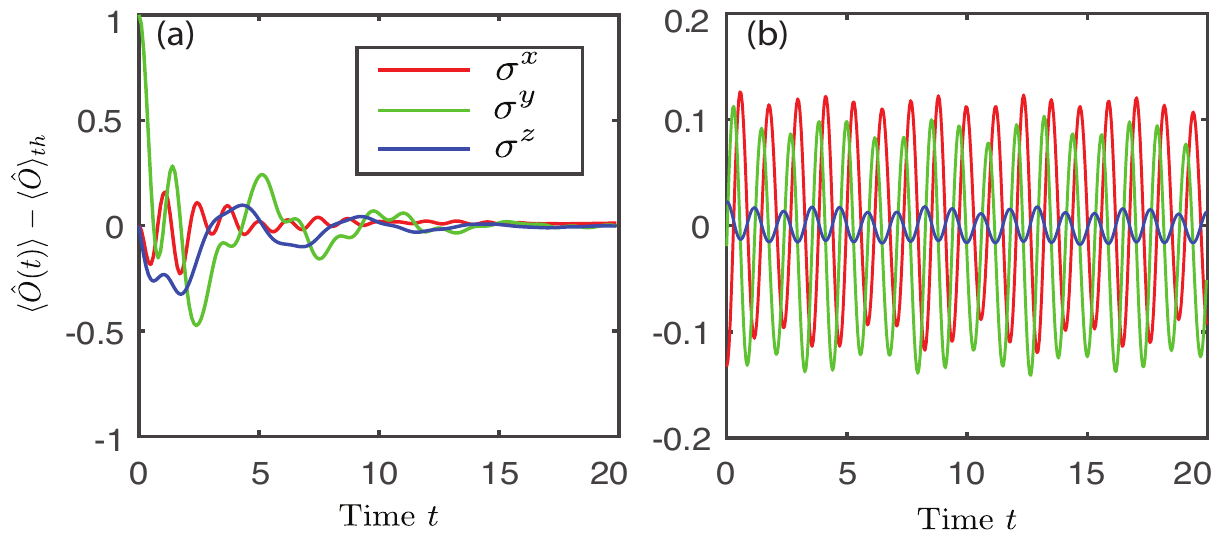}
  \caption{(color online) Strong (a) and weak (b) thermalization for different initial states.  (a) $\left\langle \sigma^{\mu}(t)\right\rangle-\left\langle \sigma^{\mu} \right\rangle _{\text{th}}$ for initial state $\ket{Y_+}$. $\left\langle \sigma^{\mu}(t)\right\rangle$ rapidly converges to its thermal value. (b) Same as (a), but for initial state $\ket{Z_+}$. The observables show strong oscillatory behavior.  Parameters: $\alpha=2.3$, $B= 0.37, L= 20$. 
}
  \label{fig4}
\end{figure}

In order to see this, we consider the time evolution of two different initial states (with the same quenched Hamiltonian): $\ket{Z_+}= \prod_j \ket{\uparrow_j}$ (the same state considered before) and $\ket{Y_+}= \prod_j \frac{1}{\sqrt{2}} (\ket{\uparrow_j}+ i\ket{\downarrow_j})$ ~\cite{Banuls11}. For $\ket{Z_+}$,   the quenched operator has the same parity as the two-kink bound state and thus the matrix elements mentioned above have non-zero values \cite{Delfino14, Delfino17}. We therefore expect slow dynamics with oscillations due to the bound quasiparticles \cite{Delfino14, Delfino17}.  On the other hand, $\ket{Y_+}$ does not satisfy this condition which suggests rapid thermalization.

We calculate the difference between the time-dependent expectation value of single-body observables, $\left\langle \sigma^{\mu}(t)\right\rangle $, and their thermal expectation value, 
$\left\langle \sigma^{\mu} \right\rangle _{\text{th}} = \text{tr}(e^{-\beta_{\Psi}H}\sigma^{\mu})/\text{tr}(e^{-\beta_{\Psi}H})$,
where the temperature, $\frac{1}{\beta_{\Psi}}$, is determined by~(see for example, \cite{Garrison15}):
\begin{equation}
\frac{\bra{\Psi} H \ket{\Psi}}{\braket{\Psi | \Psi}}=\frac{\text{tr}(He^{-\beta_{\Psi}H} )}{\text{tr}(e^{-\beta_{\Psi}H})}.
\end{equation}
Here, $\ket{\Psi}$ denotes the initial state. As illustrated in Fig.~\ref{fig4}(a), for $\ket{Y_+}$, all single-body observables converge to $\left\langle \sigma^{\mu} \right\rangle _{\text{th}}$ rapidly, indicating strong thermalization, as expected. For $\ket{Z_+}$, we instead observe strong oscillatory behavior \cite{Delfino14, Delfino17}, with Fourier frequencies consistent with the masses of the bound quasiparticles, around $\left\langle \sigma^{\mu} \right\rangle _{\text{th}}$ [Fig.~\ref{fig4}(b)]. Within the time window shown, we observe almost no decay of these observables, indicating much slower thermalization compared to $\ket{Y_+}$~\cite{Banuls11, Lin17}.

{\it Conclusions and outlook.---} We have found that the low-energy excitations of the long-range transverse-field Ising model are confined domain-walls.  These bound quasiparticles, which arise due to long-range interactions, have clear signatures in the quench dynamics of the system \cite{Kormos17, Delfino18, Delfino14, Delfino17}. Furthermore, our work shows that general quantum mechanisms of quench dynamics developed for short-range interacting systems \cite{Delfino14, Delfino17, Delfino18} hold for long-range interacting systems. These results can be readily investigated in current  trapped ion experiments \cite{Zhang17} and other AMO system with long-range interaction \cite{Saffman10,Douglas15, Bernien17}. The slow thermalization \cite{Delfino14, Delfino17} of one-point functions induced by long-range interactions has potential applications for stabilizing non-equilibrium phases of matter in the prethermal region, such as time-crystals~\cite{Zhang17Time,Hess17,Machado17} and Floquet symmetry-protected topological phases of matter~\cite{Potirniche17, Von161, Von162, Potter16, Else16}. Finally, it would be interesting to study the effects of long-range interactions on quench dynamics of $q$-state Potts models, which admit mesonic, as well as baryonic excitations~\cite{Delfino08, Lepori09, Rutke15, Lencs15}.

\begin{acknowledgments}
We are grateful to F. Verstraete, P. Calabrese, A. Bapat, J. Garrison, S.K. Chu,  C. Flower, 
and S. Whitsitt 
for useful discussions. F.L., R.L., and A.V.G.\ acknowledge support by ARO, NSF Ideas Lab on Quantum Computing, the DoE ASCR Quantum Testbed Pathfinder program, ARL CDQI, NSF PFC at JQI, and ARO MURI. 
R.L. and P.T. were supported by NIST NRC Research Postdoctoral Associateship Awards. G.P., J.Z. and C.M. are supported by the ARO and AFOSR Atomic and Molecular Physics Programs, the AFOSR MURI on Quantum Measurement and Verification, the IARPA LogiQ program, and the NSF Physics Frontier Center at JQI. This research was supported in part by the National Science Foundation under Grant No. NSF PHY-1748958.
\end{acknowledgments}

\bibliography{confinement_revise_bib}

\clearpage
\newpage 
\onecolumngrid
\setcounter{figure}{0}
\makeatletter
\renewcommand{\thefigure}{S\@arabic\c@figure}
\setcounter{equation}{0} \makeatletter
\renewcommand \theequation{S\@arabic\c@equation}
\renewcommand \thetable{S\@arabic\c@table}

\begin{center} 
{\large \bf Supplemental Material}
\end{center} 

 This Supplemental Material is organized as follows. In \cref{sec:light_spread}, we provide detailed numerical results showing light-cone spreading of correlation functions by zooming in on Figs. \ref{fig1} (b) and (c) in the main text. 
 In \cref{sec:scaling_converge}, we provide a detailed analysis on the scaling and convergence of the potential  in the main text.

\section{Light-cone spreading of correlation functions}
\label{sec:light_spread}

In the main text, we have shown that the magnitude of $\langle \sigma^z_j(t) \sigma^z_k(t)\rangle_c$ is suppressed by long-range interactions.  
{As stressed in the main text, this does not indicate the disappearance of the light-cone spreading of correlations (quantum information)}.  In this section, we provide detailed numerics showing that the light-cone behaviour is still present by zooming in on the weak-signal regions of Figs. \ref{fig1}(b) and (c) of the main text. 
\begin{figure}[!htb]
  \centering\includegraphics[width=1.0\textwidth, height=3.6cm]{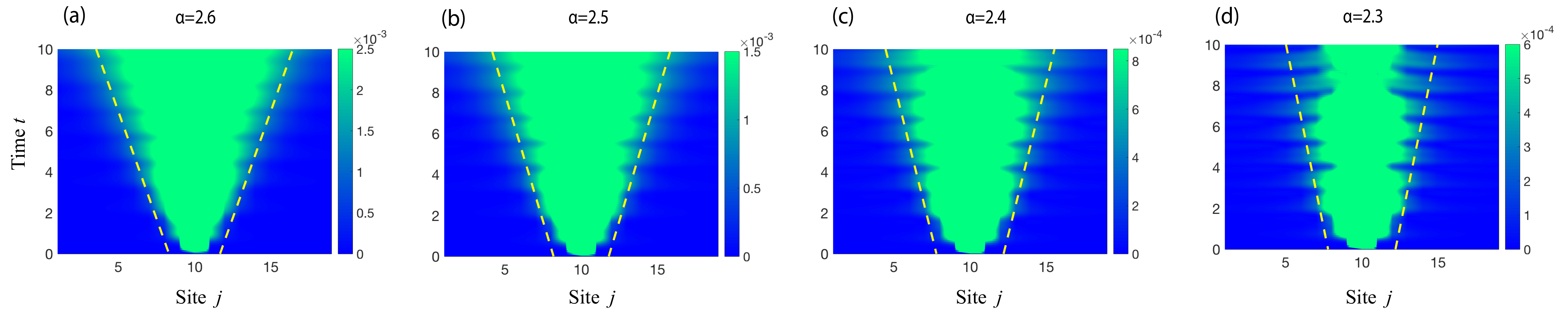}
  \caption{(color online) $ \langle \sigma^z_j \sigma^z_k\rangle_c$ after a quantum quench with initial state $|\Psi_0\rangle$. Parameters: $L= 19$, $k=10$, and $B=0.27$. (a) $\alpha= 2.6$, (b) $\alpha= 2.5$,  (c) $\alpha= 2.4$,  (d) $\alpha= 2.3$.  The green regions represent out-of-range values of the correlation functions. The  yellow dashed lines illustrate twice the maximal velocity of quasiparticles (within the three lowest energy bands). The maximal velocity for each $\alpha$ is calculated using the two-kink model, and takes the value of $v_{max}=0.24J,\   0.20J, \ 0.17J,\  0.14J $ for (a), (b), (c), and (d), respectively. The speed at which the front of the time-dependent correlation function propagates is consistent with twice the maximal velocity of the quasiparticles.  
}
  \label{fig_s1}
\end{figure}

Figs. \ref{fig_s1}(a)-(d) show correlation spreading after a sudden quench (for the same initial state, $\ket{\Psi_0}$, as in the main text) for several different $\alpha$.  Figs. \ref{fig_s1} (a) and (d)  take the same parameters of the post-quench Hamiltonians as Figs. \ref{fig1} (b) and (c) in the main text, but use an intensity scale up to two orders magnitude smaller.  By zooming in on the weak-signal regions, we observe that correlations do indeed exhibit light-cone spreading, though they may spread at different maximal velocities compared to the short-range case. These results are consistent with the general theory of quench dynamics in one-dimensional systems first formulated in Refs.\ \cite{Delfino14, Delfino17} for short-range interacting systems, where the light-cone spreading of correlations is always present with a slope equal to twice the maximal velocity of the quasiparticles. 





\section{Scaling and Convergence Analysis of  Confining Potential}
\label{sec:scaling_converge}

In this section, we provide a detailed analysis on the scaling and convergence of the potential that appears in the two-kink model. We use integrals to approximate sums. While this does not give an exact value for the potential, we will see that scaling exponents given by this approximation agree well with numerics presented in the main text. 

We use $V(n, L,\alpha)$ to denote the  potential energy of a two-domain-wall state with length $n$ on a finite chain of length $L$. The potential can be rewritten in the following form:
\begin{align}
V(n, L, \alpha)=
4 \left[ \sum_{r=1}^{L} \frac{1}{r^{\alpha}} + \sum_{r=2}^{L} \frac{1}{r^{\alpha}} +...+\sum_{r=n}^{L} \frac{1}{r^{\alpha}} -1 \right]. 
\end{align}
Note that Eq.\ (\ref{potential}) in the main text can be obtained by taking the above equation to the thermodynamic limit. 

We now approximate the above sums with integrals, which gives
\begin{align}
\widetilde{V}(n, L, \alpha) =
4 \left[ \int_1^{L} \frac{1}{r^{\alpha}} dr + \int_2^{L} \frac{1}{r^{\alpha}}dr +...+\int_n^{L} \frac{1}{r^{\alpha}}dr -1 \right]
=4 \left[   \frac{1}{\alpha -1} \left( \sum_{r=1}^n\frac{1}{r^{\alpha-1}} - \frac{n}{L^{\alpha-1}}\right) -1 \right] .
\end{align} 
After approximating the remaining sum, we obtain
\begin{equation}
\doublewidetilde{V}(n, L, \alpha) =
4 \left[   \frac{1}{\alpha -1} \left( \int_{1}^n\frac{1}{r^{\alpha-1}}dr - \frac{n}{L^{\alpha-1}}\right) -1 \right] \\
= 4 \left[ \frac{1- 1/n^{\alpha-2}}{(\alpha-1)(\alpha-2)} - \frac{n}{(\alpha-1)L^{\alpha-1}}-1 \right].
\end{equation}
Three comments are in order: (i) The second term in the above expression tells us that, for finite $n$, the potential is finite in  the thermodynamic limit ($L\rightarrow \infty$) only when $\alpha>1$. Therefore, the masses of these bound states are finite when $\alpha>1$.  This agrees with the convergence properties of the Riemann zeta function. (ii) For a finite system, the potential $\doublewidetilde{V}(n, L, \alpha)$ scales as $c_0- c_1/L^{\alpha-1}$. Since all the potential energies of the two-domain-wall states have such scaling, the masses given by eigenenergies of Eq.~\eqref{eigen} in the main text should also have the same scaling. This implies that $\beta$ (defined in the caption to Fig.~\ref{fig3} of the main text) is equal to $\alpha-1$, which is in agreement with the numerical results presented in the inset of Fig.~\ref{fig3}(d). (iii) Because of the first term of the above equation, $V(n)$ goes to infinity when $n$ goes to infinity for $1 <\alpha \leq 2$, while it is upper-bounded when $\alpha>2$. This is also reflected in Fig.~\ref{fig3}(a) of the main text. Therefore, when $\alpha>2$, the two-kink model predicts that only the lower part of the energy spectrum is composed of bound states. In other words, for a high enough energy, we have a continuum of states. However, for $\alpha \leq 2$, all eigenstates of the two-kink model are bound quasiparticles.

\end{document}